\documentclass[12pt]{article}
\usepackage{amssymb,amsmath}

\begin{document}
\title{\textbf{Stability of de Sitter Solutions in Non-local Cosmological
Models}}

\author{E. Elizalde$^{1}$\footnote{E-mail: elizalde@ieec.uab.es, elizalde@math.mit.edu} ,
E.O. Pozdeeva$^{2}$\footnote{E-mail: pozdeeva@www-hep.sinp.msu.ru} , and  S.Yu.
Vernov$^{1,2}$\footnote{E-mail: vernov@ieec.uab.es, svernov@theory.sinp.msu.ru} \vspace*{3mm} \\
\small $^1$Instituto de Ciencias del Espacio (ICE/CSIC) \, and \\
\small  Institut d'Estudis Espacials de Catalunya (IEEC) \\
\small  Campus UAB, Facultat de Ci\`encies, Torre C5-Parell-2a planta, \\
\small  E-08193, Bellaterra (Barcelona), Spain\\
\small  $^2$Skobeltsyn Institute of Nuclear Physics, Lomonosov  Moscow State University,\\
\small  Leninskie Gory 1, 119991, Moscow, Russia
}

\date{ }

\maketitle

\begin{abstract}
A non-local gravity model, which includes a function $f(\Box^{-1} R)$, where $\Box$ is the d'Alembert operator,  is considered. For the model with an exponential $f(\Box^{-1} R)$
de Sitter solutions are explored, without any
restrictions on the parameters. Using Hubble-normalized variables, the stability of the de Sitter solutions is investigated, with respect to perturbations in the Bianchi I metric, in the case of zero cosmological constant, and sufficient conditions for stability are obtained.
\end{abstract}



\section{Introduction}

Modern cosmological
observations
allow to obtain joint constraints on cosmological parameters (see, for example,~\cite{Kilbinger:2008gk}) and
indicate that the current expansion of the Universe is accelerating.
The simplest model able to reproduce this late-time cosmic
acceleration is general relativity with a cosmological constant. Other
models involve modifications of gravity, as for instance $F(R)$ gravity,
with $F(R)$ an (in principle) arbitrary function of the scalar
curvature (for reviews see~\cite{Review-Nojiri-Odintsov,Book-Capozziello-Faraoni}).

Higher-derivative corrections to the Einstein--Hilbert action are being actively studied
in the context of quantum gravity (as one of the first papers we can mention~\cite{Stelle}).
A non-local gravity theory
obtained by taking into account quantum effects has been proposed
in~\cite{Deser:2007jk}. Also, the string/M-theory is
usually considered as a possible theory for all fundamental
interactions, including gravity. The appearance of nonlocality
within string field theory is a good motivation for studying non-local
cosmological models. Most of the non-local cosmological models
explicitly include a function of the d'Alembert operator, $\Box$, and
either define a non-local modified
gravity~\cite{Non-local-gravity-Refs,Odintsov0708,Jhingan:2008ym,Non-local-FR,Koivisto:2008,Nojiri:2010pw,
Bambu1104,ZS,EPV2011} or add a non-local scalar field,
minimally coupled to gravity~\cite{Non-local_scalar}.

In this paper, we consider a modification that includes a function of
the $\Box^{-1}$ operator. This modification does not assume the
existence of a new dimensional parameter in the action and the ensuing non-local
model has a local scalar-tensor formulation. The most currently
studied example~\cite{Odintsov0708,Jhingan:2008ym,Koivisto:2008,Nojiri:2010pw,Bambu1104,ZS,EPV2011}
 of a model of this kind with de Sitter solutions is characterized by a
function $f(\Box^{-1}R)=f_0 e^{(\Box^{-1}\!R)/\beta}$, where $f_0$ and
$\beta$ are real parameters. It has been shown
in~\cite{Odintsov0708} that a theory of this kind, being consistent
with Solar System tests, may actually lead to the known Universe
history sequence: inflation, radiation/matter dominance and a dark
epoch. Expanding universe solutions $a\sim t^n$  have been found in~\cite{Odintsov0708,ZS}.
In~\cite{Koivisto:2008} the ensuing cosmology at the four basic epochs:
radiation dominated, matter dominated, accelerating, and a general
scaling has been studied for non-local models involving, in particular, an
exponential form of $f(\eta)$. An explicit mechanism to screen the cosmological constant in
non-local gravity was discussed in~\cite{Nojiri:2010pw,Bambu1104,ZS}.

De Sitter solutions play a very important role in
cosmological models, because both inflation and the late-time Universe
acceleration can be described as a de Sitter solution with
perturbations.
A few de Sitter solutions for this model have been found
in~\cite{Odintsov0708} and also analyzed in~\cite{Bambu1104}.
In~\cite{EPV2011} de Sitter solutions have been obtained without any
restriction and it has been shown that the model can have de Sitter solutions only if the
function $f(\Box^{-1}R)$ satisfies a  given second order linear differential
equation. The simplest solution of this equation is an exponential function.

The Bianchi I metric can be considered
as a minimal generalization of Friedmann--Lema\^{i}tre--Robertson--Walker (FLRW) spatially flat metric.
Considering the stability of de Sitter solutions in Bianchi I metric we include anisotropic
perturbations in our consideration.
For the model with the exponential function $f(\Box^{-1}R)$ and nonzero cosmological constant $\Lambda$ the stability of de Sitter solutions in the Bianchi I metric has been analysed in~\cite{EPV2011}.

In the case $\Lambda=0$, the stability
of the fixed point for the system of equations in terms of Hubble-normalized variables has been
discussed in \cite{Odintsov0708} and further investigated in \cite{Jhingan:2008ym,EPV2011}.
In all these papers the stability of solutions has been analysed only with respect to isotropic perturbations of the initial conditions, in other words, in the FLRW metric. Here we investigate the stability of de Sitter solutions at $\Lambda=0$ in the Bianchi I metric, and show that the stability conditions, in the Bianchi I metric and in the FLRW metric, are the same.

\section{Non-local gravitational models in the Bianchi I metric}
Consider the following action for non-local gravity
\begin{equation}
\label{nl1} S=\int d^4 x \sqrt{-g}\left\{ \frac{1}{2\kappa^2}\left[
R\Bigl(1 + f\left(\Box^{-1}R \right)\Bigr) -2 \Lambda \right] +
\mathcal{L}_\mathrm{matter} \right\} \, ,
\end{equation}
where ${\kappa}^2 \equiv 8\pi/{M_{\mathrm{Pl}}}^2$, the Planck mass
being $M_{\mathrm{Pl}}  = 1.2 \times 10^{19}$ GeV.  The determinant of the metric tensor
$g_{\mu\nu}$ is $g$,  $\Lambda$  is the
cosmological constant, $f$  is a differentiable function,  and $\mathcal{L}_\mathrm{matter}$ is the matter
Lagrangian. We use the
signature $(-,+,+,+)$.

Note that the modified gravity action (\ref{nl1}) does not include
a new dimensional parameter. This non-local model has a local
scalar-tensor formulation. Introducing two scalar fields, $\eta$ and $\xi$, we can
rewrite action~(\ref{nl1}) in the following local form:
\begin{equation}
\label{anl2} S = \int d^4 x \sqrt{-g}\left\{
\frac{1}{2\kappa^2}\left[R\left(1 + f(\eta)-\xi\right) + \xi\Box\eta  -
2 \Lambda \right] + \mathcal{L}_\mathrm{matter}  \right\} \, .
\end{equation}
By varying the action (\ref{anl2}) over $\xi$, we get $\Box\eta=R$.
Substituting $\eta=\Box^{-1}R$ into action~(\ref{anl2}), one reobtains
action~(\ref{nl1}).

Variation of  action~(\ref{anl2}) with respect to $\eta$ yields
$\Box\xi+ f'(\eta) R=0$, where the prime denotes derivative with
respect to $\eta$.
Varying action~(\ref{anl2}) with respect to the
metric tensor $g_{\mu\nu}$ yields
\begin{equation}\label{nl4}
\begin{split}
& \frac{1}{2}g_{\mu\nu} \left[R\left(1 + f(\eta) -
 \xi\right)
 - \partial_\rho \xi \partial^\rho \eta - 2 \Lambda \right]
 - R_{\mu\nu}\left(1 + f(\eta) - \xi\right)+{}  \\ &+ \frac{1}{2}\left(\partial_\mu \xi \partial_\nu \eta
+ \partial_\mu \eta \partial_\nu \xi \right)
 -\left(g_{\mu\nu}\Box - \nabla_\mu \partial_\nu\right)\left( f(\eta) -
\xi\right) + \kappa^2T_{\mathrm{matter}\, \mu\nu}=0\, ,
\end{split}
\end{equation}
where $\nabla_\mu$ is the covariant derivative and
$T_{\mathrm{matter}\,\mu\nu}$ the energy--momentum tensor of matter.

Let us
consider the Bianchi I metric with the interval
\begin{equation} \label{Bianchi}
{ds}^{2}={}-{dt}^2+a_1^2(t)dx_1^2+a_2^2(t)dx_2^2+a_3^2(t)dx_3^2.
\end{equation}
The Bianchi universe models are spatially homogeneous anisotropic
cosmological models. Interpreting the solutions of the Friedmann
equations as isotropic solutions in the Bianchi I metric, we include
anisotropic perturbations in our consideration. A similar stability
analysis has been made for cosmological models with scalar fields and
phantom scalar fields in~\cite{ABJV0903}.
It is convenient to express $a_i$ in terms of new variables $a$ and
$\beta_i$ (we use the notation of~\cite{Pereira}):
\begin{equation}
a_i(t)= a(t) e^{\beta_i(t)}.
\end{equation}
Imposing the constraint
 $\beta_1(t)+\beta_2(t)+\beta_3(t)=0$,
at any $t$, one has the following relations
\begin{equation}
a(t)=\left[a_1(t)a_2(t)a_3(t)\right]^{1/3},
\quad
H_i\equiv \frac{\dot a_i}{a_i}= H+\dot\beta_i, \quad\mbox{and}\quad
H\equiv \frac{\dot a}{a}=\frac{1}{3}(H_1+H_2+H_3).
\end{equation}
In the case of the FLRW spatially flat metric we have $a_1=a_2=a_3=a$, all $\beta_i=0$, and $H$ is the Hubble parameter.  Following~\cite{Pereira},  we
introduce the shear
\begin{equation}
\sigma^2\equiv \dot\beta_1^2+\dot\beta_2^2+\dot\beta_3^2.
\end{equation}

In the Bianchi I metric $R=12H^2+6\dot H+\sigma^2$. The equations of motion for the scalar fields are as follows:
\begin{equation}
\label{equ3} \ddot\eta={}-3H\dot\eta-12H^2-6\dot H-\sigma^2.
\end{equation}
\begin{equation}
\label{equ4}
    \ddot \xi={}-3H \dot\xi+\left(12H^2+6\dot H+\sigma^2\right)f'(\eta),
\end{equation}
For a perfect matter fluid, we have $T_{\mathrm{matter}\, 0 0} =
\rho_{\mathrm{m}}$ and $T_{\mathrm{matter}\, i j} = P_{\mathrm{m}} g_{i
j}$. The  equation of state  is
\begin{equation}
\label{equ_rho} \dot\rho_{\mathrm{m}}={}-
3H(P_{\mathrm{m}}+\rho_{\mathrm{m}}).
\end{equation}
The Einstein equations have the form:
\begin{equation}
\label{equ1B} {}\left[\frac{\sigma^2}{2}- 3 H^2\right]\!\left(1 +  \phi
- \xi\right) + \frac{1}{2}\dot\xi \dot\eta
 - 3H\left( \dot\phi -\dot\xi \right)  + \Lambda
+ \kappa^2 \rho_{\mathrm{m}}=0\, ,
\end{equation}
\begin{equation}
\label{equ2B} \left[2\dot H +
3H^2+\frac{\sigma^2}{2}-\ddot\beta_j-3H\dot\beta_j\right]\! \left(1 +
 \phi - \xi\right) + \frac{1}{2}\dot\xi\dot\eta + \ddot\phi -\ddot\xi +
(2H-\dot\beta_j)(\dot\phi -\dot\xi)  = \Lambda - \kappa^2
P_{\mathrm{m}},
\end{equation}
where $\phi\equiv f(\eta)$. Summing Eqs.~(\ref{equ2B}) for $j=1,2,3$, we
get
\begin{equation}
\label{equ2Bsum} \left[2\dot H + 3H^2+\frac{\sigma^2}{2}\right]\!
\left(1 + \phi - \xi\right) + \frac{1}{2}\dot\xi \dot\eta +
\ddot\phi-\ddot\xi + 2H\left(\dot\phi -\dot\xi\right)  = \Lambda -
\kappa^2 P_{\mathrm{m}}.
\end{equation}
From equations (\ref{equ2B}) it is easy to get
\begin{equation}
\left[\ddot\beta_j+3H\dot\beta_j\right]\! \left(1 + \phi -
\xi\right)+\dot\beta_j\left(\dot\phi -\dot\xi\right)=0,
\label{equbeta}
\end{equation}
\begin{equation}
\label{equvartheta}
\left[\frac{d}{dt}\left(\sigma^2\right)+6H\sigma^2\right]\! \left(1 +
\phi - \xi\right)+2\sigma^2\left(\dot\phi -\dot\xi\right)=0.
\end{equation}
The functions $H(t)$, $\sigma^2(t)$,  $\xi(t)$,
$\eta(t)$, and $\rho_m(t)$ can be obtained from equations
 (\ref{equ3})--(\ref{equ1B}), (\ref{equ2Bsum}) and
(\ref{equvartheta}). If $H(t)$ and the scalar fields are known, then
$\beta_j(t)$ can be found from (\ref{equbeta}).

Following~\cite{Bambu1104}, we consider matter with a state
parameter $w_{\mathrm{m}}\equiv P_{\mathrm{m}}/\rho_{\mathrm{m}}$ to be a
constant not equal to $-1$. Thus, Eq.~(\ref{equ_rho}) has the
following general solution:
\begin{equation}
\rho_{\mathrm{m}}=\rho_0\,e^{{}-3(1+w_{\mathrm{m}})H_0t},
\end{equation}
where $\rho_0$ is an arbitrary constant.

 It has been shown in~\cite{EPV2011} that the model (\ref{anl2}) can have de Sitter solutions
for functions $f$ of the following forms:
\begin{equation}
\label{f1}
 f_1(\eta)
=\frac{C_2}{4}e^{\eta/2}+C_3e^{3\eta/2}+C_4-\frac{\kappa^2\rho_0}{3(1+3w_{\mathrm{m}})H_0^2}
e^{3(w_{\mathrm{m}}+1)\eta/4}\,,\quad \mbox{for}\quad w_{\mathrm{m}}\neq {}-\frac{1}{3}\,,
\end{equation}
\begin{equation}
\label{fw13} \tilde{f}_1(\eta)
=\frac{C_2}{4}e^{\eta/2}+C_3e^{3\eta/2}+C_4+\frac{\kappa^2\rho_0}{4H_0^2}\left(1-\frac{1}{3}\eta\right)
e^{\eta/2}, \quad \mbox{for}\quad w_{\mathrm{m}}= {}-\frac{1}{3}\,,
\end{equation}
where $C_2$, $C_3$, and $C_4$ are arbitrary constants.
One can see that the key ingredient common to all these functions $f_i(\eta)$ is the
exponential form. For the models with $f(\eta)$ equal to a simple exponential function or to a sum of exponential functions, particular de Sitter solutions have been found
in~\cite{Odintsov0708,Bambu1104}. In the most general form, de Sitter solutions for the case of the exponential function $f(\eta)$ have been obtained in~\cite{EPV2011}.

\section{De Sitter solutions and their stability}

Let us consider the action (\ref{anl2}),  with
\begin{equation}
\label{f} f(\eta)=f_0 e^{\eta/{\beta}}\, ,
\end{equation}
where $f_0$ and $\beta$ are real constants. This form of $f(\eta)$ is the
simplest function which belongs to the set of functions described by
(\ref{f1}). De Sitter solutions with a constant nonzero $H=H_0$
have the following expression~\cite{EPV2011}
\begin{eqnarray}
\label{eta} \eta(t) &=& {}-4H_0(t-t_0),\\
\label{sol_eta}
    \xi(t)&=&{}-\frac{3f_0\beta}{3\beta-4}e^{-4H_0(t-t_0)/\beta}+\frac{c_0}{3H_0}e^{-3H_0(t-t_0)}-\xi_0,
    \quad \mbox{at}\quad \beta\neq 4/3,\label{sol_xi}\\
    \xi(t) &=& {}-f_0(c_0+3H_0(t-t_0))e^{-3H_0(t-t_0)}- \xi_0, \quad \mbox{at}\quad \beta= 4/3,
\end{eqnarray}
where $c_0$ and $t_0$ are arbitrary constants,
\begin{equation}
\xi_0 ={}-1 -\frac{\Lambda}{3H_0^2},\qquad \rho_{0} = \frac{6
\left(\beta-2\right) H_0^2f_0}{\kappa^2\beta}\ ,\qquad w_{\mathrm{m}} =
{}-1+\frac{4}{3\beta}.
\label{CondeSit}
\end{equation}

The case $\beta=2$ corresponds to $\rho_0=0$. Thus, the model with
exponential $f(\eta)$ has no de Sitter solution if we add matter with $w_{\mathrm{m}} =-1/3$.
The type of function $f(\eta)$, which can have such solutions,
is given by (\ref{fw13}).
The case $\beta=4/3$ corresponds to dark matter, because
$w_{\mathrm{m}}=0$.

Using (\ref{equ3}) and (\ref{equ4}), we get equation~(\ref{equ2Bsum}) in the  form
\begin{equation}
\label{equ12n2B}
 2\left[1 + \frac{\beta-6}{\beta}\phi-
\xi\right]\dot H =4H\left[\frac{\phi\dot\eta}{\beta}-\dot\xi\right]-\frac{\phi\dot\eta^2}{\beta^2}
+\frac{24}{\beta}H^2\phi - \dot\xi \dot\eta - \frac{4\kappa^2}{3\beta}
\rho_{\mathrm{m}}-\left[1 + \frac{\beta-2}{\beta}\phi-
\xi\right]\sigma^2.
\end{equation}
For $H_0>0$ and $\beta>0$,
\begin{equation}
\phi\rightarrow 0,\qquad\xi\rightarrow {}-\xi_0, \quad\mbox{at}\quad t \rightarrow +\infty.
\end{equation}
Therefore, the coefficient of $\dot H$ in (\ref{equ12n2B}) tends to
$\Lambda/(3H_0^2)$.  In the case of nonzero $\Lambda$, the stability of de Sitter solutions at late times can be analysed
without using of the Hubble-normalized variables. It has been found in~\cite{EPV2011} that for  $H_0>0$ and $\beta>0$, the de Sitter solutions are stable with respect to fluctuations of the initial conditions in the Bianchi I metric  at any nonzero value of $\Lambda$.

Here we consider the stability of de Sitter solutions
with respect to fluctuations of the initial conditions in the Bianchi I metric, in the case $\Lambda=0$.
To analyze the stability of the de Sitter solutions at $\Lambda=0$,
we transform the system of equations
using the Hubble-normalized variables
\begin{equation}
X={}-\frac{\dot{\eta}}{4H},\qquad
  W= \frac{\dot{\xi}}{6Hf},\qquad
  Y=\frac{1-\xi}{3f},\qquad
  Z=\frac{\kappa^2\rho_m}{3H^2f}, \qquad  K=\frac{\sigma^2}{2H^2}
\end{equation}
and the independent variable, $N$,
\begin{equation}
\frac{d}{dN}\equiv a\frac{d}{da}= \frac{1}{H}\frac{d}{dt}\ .
\end{equation}
The use of these variables makes the equation of motion dimensionless.
 Equations~(\ref{equ3}), (\ref{equ4}), (\ref{equ_rho}), and (\ref{equvartheta}) are equivalent to the following
ones, in terms of the new variables,
\begin{eqnarray}
 \frac{dX}{dN}&=&3(1-X)+\frac{1}{H}\left(\frac32-X\right)\frac{dH}{dN}+\frac{K}{2}\,,\label{dXsys} \\
 \frac{dW}{dN}&=&\frac{2}{\beta}\left(1+2WX\right)-3W+\frac{1}{H}\left(\frac{1}{\beta}
 -W\right)\frac{dH}{dN}+\frac{K}{3\beta}\,,\label{dWsys}\\
 \frac{dZ}{dN}&=&\frac{4}{\beta}(X-1)Z-2\frac{Z}{H}\frac{dH}{dN}\,,\label{dZsys}\\
 \left(\frac{dK}{dN}\right.&+&\left.\frac{2K}{H}\frac{dH}{dN}+6K\right)(3Y+1)=4K\left(\frac{2X}{\beta}+3W\right)\label{eq5}.
\end{eqnarray}

To get the full system of the first order differential equations we need to get one for $\frac{dH}{dN}$ and to eliminate
$Y$. To do this, we use Eq.~(\ref{equ1B}), which  can be written in
terms of the new variables as
\begin{equation}
Y={}-\frac{1}{3}+\frac{2\beta(2X-3)W-4X-\beta Z}{\beta(K-3)}. \label{equ1nv}
\end{equation}
 Differentiating (\ref{equ1nv}), substituting (\ref{dXsys})--(\ref{eq5}),  and using
\begin{equation}
\frac{dY}{dN}=2\left(\frac{2XY}{\beta}-W\right)=\frac{4X}{3\beta^2(3-K)}\Bigl(\beta(K-3)+6\beta(3-2X)W+12X+3\beta Z\Bigr)-2W,\label{dY}
\end{equation}
one gets
\begin{equation}
\label{dH}
\begin{split}
\Bigl(2(2X-3)(\beta W-1)-\beta Z-2K\Bigr)\frac{1}{H}\frac{dH}{dN}&=\frac{8(3-K)X^2}{3\beta}+{}\\
{}+\frac{4}{3}(6-9\beta W+K)X+2Z+12(\beta W-1)&+\left(2-\frac{2}{3}Z+(2W+Z)\beta\right)K+\frac{2}{3}K^2,
\end{split}
\end{equation}

In terms of the new variables, the system (\ref{dXsys})--(\ref{eq5}), (\ref{dH}) has the following fixed point
\begin{equation}
\label{FixPoi}
H=H_0,\qquad  X_0=1,\qquad Z_0=\frac{2(\beta-2)}{\beta}, \qquad W_0=\frac{2}{3\beta-4}, \qquad K_0=0,
\end{equation}
which corresponds to de Sitter solution for $\beta\neq 4/3$, with $c_{0}=0$.
In the case of an arbitrary $c_0$, for the de Sitter
solution, we get
\begin{equation}
W=\frac{2}{3\beta-4}-\frac{c_0}{6H_0f_0}e^{-(3-4/\beta)(N-N_0)},
\end{equation}
where $N_0=H_0t_0$. The function $W$ tends to infinity at large $N$ for $\beta<4/3$ and
$\lim\limits_{N\rightarrow\infty}  W=W_0$ at $\beta>4/3$. So, the fixed point can be stable
only at $\beta>4/3$.  Under this condition all de Sitter solutions tend to a fixed point, what means that,
for any $\varepsilon>0$, there exists a number, $N_1$, such that the de Sitter
solution is in the $\varepsilon/2$ neighborhood of the fixed point, for all $N>N_1$.
Therefore, the stability of the fixed point guarantees the stability of all
de Sitter solutions.

For $\beta=4/3$ the function $W$,
corresponding to de Sitter solutions, depends on $N$ for any value of parameters. Thus,
this choice of dimensionless variable is not suitable to analyse
stability of the de Sitter solutions for  $\beta=4/3$. Here we will deal with the case $\beta\neq 4/3$, only.

Let us consider perturbations in the neighborhood of (\ref{FixPoi}):
\begin{equation}
X=1+\varepsilon x_1,\quad Z=Z_0(1+\varepsilon z_1),\quad
W=W_0(1+\varepsilon w_1),\quad H=H_0(1+\varepsilon h_1),\quad K=\varepsilon k_1,
\end{equation}
where $\varepsilon$ is a small parameter.
To first order in $\varepsilon$, after some work we obtain the system of linear
equations:
\begin{equation}
\label{equx1z1} \frac{dx_1}{dN}={}-3x_1+\frac{1}{2}\frac{dh_1}{dN}+\frac{1}{2} k_1,\qquad  \frac{dz_1}{dN}=\frac{4}{\beta}x_1-2\frac{dh_1}{dN},
\end{equation}
\begin{equation}
\label{equw1}
\frac{dw_1}{dN}=\frac{4}{\beta}x_1+\frac{\beta-4}{2\beta}\frac{dh_1}{dN}+\left(\frac{4}{\beta}-3\right)w_1
+\frac{3\beta-4}{6\beta}k_1,
\end{equation}
\begin{equation}
\label{equh1}
\frac{dh_1}{dN}=\frac{8(4-\beta)}{\beta(3\beta^2-11\beta+12)}x_1
-\frac{2(3\beta-4)(\beta-2)}{\beta(3\beta^2-11\beta+12)}z_1-\frac{3\beta^2-5\beta+4}{3\beta^2-11\beta+12} k_1,
\end{equation}
\begin{equation}
\label{equk1}
\frac{dk_1}{dN}=\left(\frac{8}\beta-6\right)k_1.
\end{equation}
Solving (\ref{equk1}), we get
\begin{equation}
k_1(N)=b_1e^{-(6-8/\beta)N},
\end{equation}
where $b_1$ is an arbitrary constant and $k_1$ tends to zero for $\displaystyle N\rightarrow\infty$, if and only if  $\beta>4/3$.

Substituting $k_1$ and (\ref{equh1}) into (\ref{equx1z1}), we get a system of two inhomogeneous differential equations. As known, the general  solution of this system is a sum of the general solution of the corresponding homogeneous system and a particular solution of inhomogeneous one. The homogeneous system corresponds to the FLRW metric (the case $K=0$) and those general solution, which has been obtained in~\cite{EPV2011}, is bounded and tends to zero for $N\rightarrow\infty$, if $4/3<\beta\leqslant 2$.
For any $\beta$ from this interval a particular solution of the inhomogeneous system tends to zero as well, because
$k_1$ tends to zero at $\beta>4/3$. Therefore, the perturbations $x_1$ and $z_1$ decrease provided $4/3<\beta<2$.
Substituting $x_1(N)$ and $z_1(N)$ into Eqs.~(\ref{equw1}) and
(\ref{equh1}) we get that $h_1(N)$ and $w_1(N)$ decrease as well. Note
that $h_1(N)$ has a part, $H_{1}$, which does not depend on $N$ and,
therefore, it can be considered as part of $H_0$. This result corresponds
to the fact that, for $\Lambda=0$, the value of $H_0$ can be selected
arbitrarily; thus, one can choose $\tilde{H}_0=H_0+H_{1}$ instead of
$H_0$. We can summarize the above saying
that the de Sitter solutions are stable with respect to perturbations
of the Bianchi I metric, in the case $4/3<\beta\leqslant 2$. If $f_0>0$, then the stable de Sitter
solution corresponds to $\rho_0\leqslant 0$.

\section{Conclusions}

We have  investigated de Sitter solutions in the non-local
gravity model  described by the action (\ref{nl1}) (see~\cite{Odintsov0708}). We have used
the local formulation of the model (\ref{anl2}), which includes two
scalar fields. We have specifically considered the case of the exponential function
$f(\eta)$, which is the simplest and most studied case, corresponding to the model (\ref{anl2}),
that admits de Sitter solutions.

In~\cite{EPV2011}, we have discussed the stability of de Sitter solutions in the Bianchi I metrics and obtained
that, for  $H_0>0$ and $\beta>0$, de Sitter solutions are stable, for all
nonzero values of $\Lambda$. Here we have proved that in the case $\Lambda=0$ de Sitter solutions
are stable for $H_0>0$ and $4/3<\beta\leqslant 2$. Thus, our conclusion is that de Sitter solutions, which are stable with respect to isotropic perturbations, are also stable with respect to anisotropic perturbations of the Bianchi I metric.
\medskip

The authors thank Sergei~D.~Odintsov
for useful discussions. E.E. was supported in
part by MICINN (Spain), projects FIS2006-02842 and FIS2010-15640, by
the CPAN Consolider Ingenio Project, and by AGAUR (Generalitat de
Ca\-ta\-lu\-nya), contract 2009SGR-994. E.P. and S.V. are supported in
part by the RFBR grant 11-01-00894, E.P. also by a state
contract of the Russian Ministry of Education and Science
14.740.12.0846, and S.V. by a grant of the Russian Ministry of
Education and Science NSh-4142.2010.2 and by CPAN10-PD12 (ICE,
Barcelona, Spain).


\begin{thebibliography}{99}

\bibitem{Kilbinger:2008gk}
  M.~Kilbinger {\it et al.},
\textit{Dark energy constraints and correlations with
systematics from CFHTLS weak lensing, SNLS supernovae Ia and WMAP5},
\textit{Astron. Astrophys.} \textbf{497} (2009) 677--688  [\texttt{arXiv:0810.5129}]


\bibitem{Review-Nojiri-Odintsov}
  S.~Nojiri and S.D.~Odintsov,
  \textit{Unified cosmic history in modified gravity: from F(R) theory to Lorentz
  non-invariant models},
  \textit{Phys. Rept.} \textbf{505} (2011) 59--144 [\texttt{arXiv:1011.0544}]



\bibitem{Book-Capozziello-Faraoni}
S.~Capozziello and V.~Faraoni, \textit{Beyond Einstein Gravity: A
Survey of Gravitational Theories for Cosmology and Astrophysics}, Fund.
Theor. Phys. \textbf{170}, Springer, New York, 2011

\bibitem{Stelle}
K.S.~Stelle,
  \textit{Renormalization of Higher Derivative Quantum Gravity},
  \textit{Phys. Rev. D} \textbf{16} (1977) 953--969

\bibitem{Deser:2007jk}
  S.~Deser and R.P.~Woodard,
  \textit{Nonlocal Cosmology,}
  \textit{Phys. Rev. Lett.}  {\bf 99} (2007) 111301
  [\texttt{arXiv:0706.2151}]



\bibitem{Non-local-gravity-Refs}
 T. Biswas, A. Mazumdar, and W. Siegel,
 \emph{Bouncing Universes in String-inspired Gravity},
\textit{JCAP}  \textbf{0603} (2006)  009 [\texttt{hep-th/0508194}]; \\
  S.~Capozziello, E.~Elizalde, S.~Nojiri, and S.D.~Odintsov,
  \textit{Accelerating cosmologies from non-local higher-derivative gravity},
  \textit{Phys. Lett.}  B {\bf 671}  (2009)
 193--198  [\texttt{arXiv:0809.1535}]; \\
  S.~Nesseris and A.~Mazumdar,
  \textit{Newton's constant in $f(R,R_{\mu\nu}R^{\mu\nu},\Box R)$ theories of gravity
  and constraints from BBN},  \textit{Phys.\ Rev.}  D  \textbf{79} (2009) 104006
  [\texttt{arXiv:0902.1185}]; \\
  C.~Deffayet and R.P.~Woodard,
  \textit{Reconstructing the Distortion Function for Nonlocal Cosmology},
  \textit{JCAP} {\bf 0908} (2009) 023 [\texttt{arXiv:0904.0961}]; \\
  G.~Cognola, E.~Elizalde, S.~Nojiri, S.D.~Odintsov, and S.~Zerbini,
  \textit{One-loop effective action for non-local modified Gauss-Bonnet gravity
  in de  Sitter space},
 \textit{ Eur.\ Phys.\ J.}  C {\bf 64} (2009) 483--494
  [\texttt{arXiv:0905.0543}]; \\
  G.~Calcagni and G.~Nardelli,
  \textit{Nonlocal gravity and the diffusion equation},  \textit{Phys. Rev.} D {\bf  82} (2010) 123518
  [\texttt{arXiv:1004.5144}]; \\
T.~Biswas, T.~Koivisto, and A.~Mazumdar,
  \textit{Towards a resolution of the cosmological singularity in non-local higher
  derivative theories of gravity},
  \textit{JCAP} {\bf 1011} (2010) 008 [\texttt{arXiv:1005.0590}]; \\
A.O.~Barvinsky,
\textit{Nonlocal gravity and its cosmological manifestations},
\texttt{arXiv:1107.1463}

\bibitem{Odintsov0708}
  S.~Nojiri and S.D.~Odintsov,
  \textit{Modified non-local-F(R) gravity as the key for the inflation and dark
  energy},
  \textit{Phys.\ Lett.}  B {\bf 659} (2008)
  821  [\texttt{arXiv:0708.0924}]


\bibitem{Jhingan:2008ym}
  S.~Jhingan, S.~Nojiri, S.D.~Odintsov, M.~Sami, I.~Thongkool, and S.~Zerbini,
  \textit{Phantom and non-phantom dark energy: The cosmological relevance of
  non-locally corrected gravity},
  \textit{Phys. Lett.}  B {\bf 663} (2008) 424--428 [\texttt{arXiv:0803.2613}]

\bibitem{Non-local-FR}
 K.A.~Bronnikov and E.~Elizalde,
  \textit{Spherical systems in models of nonlocally corrected gravity},
  \textit{Phys.\ Rev.}  D {\bf 81} (2010) 044032
  [\texttt{arXiv:0910.3929}]; \\
J.~Kluson,
 \textit{Non-Local Gravity from Hamiltonian Point of View},
 \textit{JHEP} {\bf 1109} (2011) 001 [\texttt{arXiv:1105.6056}]


\bibitem{Koivisto:2008}
T.S.~Koivisto,
  \textit{Dynamics of Nonlocal Cosmology},
  \textit{Phys.\ Rev.}  D {\bf 77} (2008) 123513
  [\texttt{arXiv:0803.3399}]

\bibitem{Nojiri:2010pw}
  S.~Nojiri, S.D.~Odintsov, M.~Sasaki and Y.l.~Zhang,
  \textit{Screening of cosmological constant in non-local gravity},
  \textit{Phys.\ Lett.}  B {\bf 696} (2011) 278--282
  [\texttt{arXiv:1010.5375}]

\bibitem{Bambu1104}
K. Bamba, Sh. Nojiri, S.D. Odintsov, and M. Sasaki,
\textit{Screening of
cosmological constant for De Sitter Universe in non-local gravity,
phantom-divide crossing and finite-time future singularities},
\texttt{arXiv:1104.2692}

\bibitem{ZS}
Y.l.~Zhang and M.~Sasaki, \textit{Screening of cosmological constant in non-local cosmology},
\emph{Int. J. Mod. Phys.} D \textbf{21} (2012) 1250006 [\texttt{arXiv:1108.2112}]

\bibitem{EPV2011}
 E. Elizalde, E.O. Pozdeeva, and S.Yu. Vernov,
\textit{De Sitter Universe in Non-local Gravity},
\textit{Phys. Rev.} D 2012, to be published, \texttt{arXiv:1110.5806}

\bibitem{Non-local_scalar}
G. Calcagni,
\textit{Cosmological tachyon from cubic string field
theory},
 \textit{JHEP} \textbf{0605} (2006) 012  [\texttt{hep-th/0512259}]; \\
A.S. Koshelev,
 \textit{Non-local SFT Tachyon and Cosmology},
 \textit{JHEP}
\textbf{0704} (2007) 029  [\texttt{hep-th/0701103}]; \\
I.Ya. Aref'eva, L.V. Joukovskaya, and S.Yu. Vernov,
 \textit{Bouncing and accelerating solutions in nonlocal stringy models},
 \textit{JHEP} \textbf{0707} (2007) 087 [\texttt{hep-th/0701184}]; \\
 I.Ya. Aref'eva and  I.V.  Volovich,
 \textit{Quantization of the Riemann
 Zeta-Function and Cosmology},
 \textit{Int. J. of Geom. Meth.
Mod. Phys.}  \textbf{4} (2007) 881--895 [\texttt{hep-th/0701284}]; \\
  G. Calcagni, M. Montobbio, and  G. Nardelli,
\textit{A route to nonlocal cosmology},
\textit{Phys. Rev.} D \textbf{76} (2007) 126001
[\texttt{arXiv:0705.3043}]; \\
N. Barnaby, T. Biswas, and  J.M. Cline,
\textit{p-adic Inflation},
\textit{JHEP} \textbf{0704} (2007) 056
 [\texttt{hep-th/0612230}]; \\
 I.Ya. Aref'eva, L.V. Joukovskaya, and S.Yu. Vernov,
 \textit{Dynamics in nonlocal linear models in the Friedmann--Robertson--Walker metric},
 \textit{J. Phys. A: Math. Theor.} \textbf{41} (2008) 304003
[\texttt{arXiv:0711.1364}]; \\
D.J. Mulryne and N.J. Nunes,
\textit{Diffusing non-local inflation:
Solving the field equations as an initial value problem},
\textit{Phys. Rev.}  D \textbf{78} (2008) 063519 [\texttt{arXiv:0805.0449}]; \\
 G. Calcagni, M. Montobbio, and  G. Nardelli,
\textit{Localization of nonlocal theories},
\textit{Phys. Lett.} B \textbf{662} (2008) 285--289 [\texttt{arXiv:0712.2237}]; \\
L.V. Joukovskaya,
{\it Dynamics in nonlocal cosmological models derived from string field theory},
  \textit{Phys. Rev.} D \textbf{76} (2007) 105007 [\texttt{arXiv:0707.1545}]; \\
N. Barnaby and  N. Kamran,
\textit{Dynamics with Infinitely Many
Derivatives: The Initial Value Problem},
\textit{JHEP} {\bf 0802} (2008) 008 [\texttt{arXiv:0709.3968}];\\
S.Yu. Vernov,
\textit{Localization of nonlocal cosmological models
with quadratic potentials in the case of double roots},
\textit{Class. Quant. Grav.} \textbf{27} (2010) 035006  [\texttt{arXiv:0907.0468}]; \\
S.Yu.~Vernov,
 \textit{Localization of the SFT inspired Nonlocal Linear Models and Exact
  Solutions},
\textit{Phys. Part. Nucl. Lett.} \textbf{8} (2011) 310--320  [\texttt{arXiv:1005.0372}]; \\
A.S. Koshelev and S.Yu. Vernov,
\textit{Analysis of scalar perturbations
  in cosmological models with a non-local scalar field},
\textit{Class. Quant. Grav.} \textbf{28} (2011) 085019 [\texttt{arXiv:1009.0746}]

\bibitem{ABJV0903}
I.Ya. Aref'eva, N.V. Bulatov, L.V. Joukovskaya,  S.Yu. Vernov,
\textit{Null Energy Condition Violation and Classical Stability in the
Bianchi I Metric},
\textit{Phys. Rev.} D \textbf{80} (2009) 083532 [\texttt{arXiv:0903.5264}];\\
I.Ya. Aref'eva, N.V. Bulatov, and S.Yu. Vernov,
\textit{Stable Exact Solutions in Cosmological Models with Two Scalar Fields,}
\textit{Theor. Math. Phys.} \textbf{163} (2010) 788--803 [\texttt{arXiv:0911.5105}]


\bibitem{Pereira}
T.S. Pereira, C. Pitrou, and J.-Ph. Uzan,
\textit{Theory of cosmological
perturbations in an anisotropic universe},
\textit{JCAP} \textbf{0709} (2007) 006 [\texttt{arXiv:0707.0736}]


\end{thebibliography}
\end{document}